\documentclass[12pt]{iopart}

\usepackage{bm}
\usepackage{amssymb}
\usepackage{iopams}
\usepackage{latexsym}
\usepackage{exscale}

\newcommand{\vect}[1]{\bm{#1}}
\newcommand{\ten}[1]{\mbox{\textbf{\textsf{#1}}}}

\newcommand{\tenszero}{\mbox{\textbf{\textsf{0}}}}
\newcommand{\sprod}{\!\cdot\!}
\newcommand{\tprod}{}

\newcommand{\trace}{\Tr}
\newcommand{\trans}{\mathsf{T}}
\newcommand{\dif}{\rmd}
\newcommand{\mi}{\rmi} 
\newcommand{\me}{\rme}

\begin{document}

\title{Nonequilibrium thermal Casimir--Polder forces}

\author{S Y Buhmann and S Scheel}

\address{Quantum Optics and Laser Science, Blackett Laboratory,
Imperial College London, Prince Consort Road,
London SW7 2BW, United Kingdom}
\ead{s.buhmann@imperial.ac.uk}

\begin{abstract}
We study the nonequilibrium Casimir--Polder force on an atom prepared
in an incoherent superposition of internal energy-eigenstates, which
is placed in a magnetoelectric environment of nonuniform temperature.
After solving the coupled atom--field dynamics within the framework of
macroscopic quantum electrodynamics, we derive a general expression
for the thermal Casimir--Polder force. 
\end{abstract}



\section{Introduction}
\label{sec1}

The thermal fluctuations of the electromagnetic field present at
finite temperature may interact with a single atom or molecule,
resulting in the thermal Casimir--Polder (CP) force. It has been
studied theoretically via linear response theory \cite{0037}, Lifshitz
theory \cite{0057}, quantum electrodynamics (QED) \cite{0034} as well
as heuristic generalisations of the zero-temperature result
\cite{0359} and experimentally by spectroscopic means \cite{0145}.

Thermal CP forces in nonequilibrium scenarios have recently received
an increased attention, where two cases can be distinguished. A
nonequilibrium environment can be realised when an atom interacts
with a body whose temperature differs from the environment temperature
\cite{0399}. Alternatively, situations have been considered where the
environment is at global thermal equilibrium with a uniform
temperature, but one or two present atoms are not fully thermalised
with their local environment \cite{0824,Sherkunov08}. In the present
paper, we generalise these two special cases to the full
nonequilibrium situation of an atom in a nonequilibrium state placed
within an arbitrary magnetoelectric environment of nonuniform
temperature. We use macroscopic QED (recalled in Sec.~\ref{sec2}) to
calculate the CP force from the thermal average of the quantum Lorentz
force (Sec.~\ref{sec3}) followed by a short summary
(Sec.~\ref{sec4}). 


\section{Macroscopic quantum electrodynamics at finite temperature}
\label{sec2}

Consider an environment of dispersive and absorbing magnetoelectrics
of (relative) electric permittivity $\varepsilon(\vect{r},\omega)$ and
a magnetic permeability $\mu(\vect{r},\omega)$, which are both
satisfying the Kramers--Kronig relations. This environment together
with its electromagnetic field can be characterised by a Hamiltonian
$\hat{H}_F\!=\!\sum_{\lambda=e,m}\int\dif^3r \int_0^\infty
 \dif\omega\,\hbar\omega\,
 \hat{\vect{f}}_\lambda^\dagger(\vect{r},\omega)
 \sprod\hat{\vect{f}}_\lambda(\vect{r},\omega)$ \cite{0002,0696},
where the bosonic variables $\hat{\vect{f}}_\lambda$ and
$\hat{\vect{f}}_\lambda^\dagger$ are associated with the noise
polarisation ($\lambda\!=\!e$) and magnetisation ($\lambda\!=\!m$) of
the media. When a single atom or molecule with Hamiltonian
$\hat{H}_A\!=\!\sum_n E_n |n\rangle\langle n|$ ($E_n$: eigenenergies,
$|n\rangle$: molecular eigenstates) is placed at position
$\vect{r}_{\!A}$ in this environment, it interacts with the
electromagnetic field via an electric-dipole Hamiltonian
$\hat{H}_{AF}\!=\!-\sum_{m,n}\vect{d}_{mn}\sprod
 \hat{\vect{E}}(\vect{r}_{\!A})\hat{A}_{mn}$ ($\vect{d}_{mn}$ $\!=$
$\!\langle m|\hat{\vect{d}}|n\rangle$, $\hat{A}_{mn}$
$\!=$ $\!|m\rangle\langle n|$), so that the total Hamiltonian of the
system reads $\hat{H}\!=\!\hat{H}_A+\hat{H}_F+\hat{H}_{AF}$. 

The electric field can be expressed in terms of the dynamical
variables according to
\begin{equation}
\label{1}
 \fl\hat{\vect{E}}(\vect{r})
=\int_0^{\infty}\!\dif\omega\,
 \underline{\hat{\vect{E}}}(\vect{r},\omega)+\mathrm{H.c.}
=\int_0^{\infty}\!\dif\omega\,\sum_{\lambda={e},{m}}
 \int\dif^3r'\,\ten{G}_\lambda(\vect{r},\vect{r}',\omega)
 \!\cdot\!\hat{\vect{f}}_\lambda(\vect{r}',\omega)+\mathrm{H.c.},
\end{equation}
with the coefficients $\ten{G}_\lambda$ being related
to the classical Green tensor, $\ten{G}$, by
\begin{eqnarray}
\label{2}
\ten{G}_e(\vect{r},\vect{r}',\omega)
 =\mi\,\frac{\omega^2}{c^2}
 \sqrt{\frac{\hbar}{\pi\varepsilon_0}\,
 \mathrm{Im}\varepsilon(\vect{r}',\omega)}\,
 \ten{G}(\vect{r},\vect{r}',\omega),\\[1ex]
\label{3}
\ten{G}_m(\vect{r},\vect{r}',\omega)
 =\mi\,\frac{\omega}{c}
 \sqrt{\frac{\hbar}{\pi\varepsilon_0}\,
 \frac{\mathrm{Im}\mu(\vect{r}',\omega)}
 {|\mu(\vect{r}',\omega)|^2}}
 \bigl[\vect{\nabla}'
 \!\!\times\!\ten{G}(\vect{r}',\vect{r},\omega)
 \bigr]^{\trans}.
\end{eqnarray}
The Green tensor is the unique solution to the Helmholtz equation
\begin{equation}
\label{4}
\left[\bm{\nabla}\times
\frac{1}{\mu(\vect{r},\omega)}\bm{\nabla}\times
 \,-\,\frac{\omega^2}{c^2}\,\varepsilon(\vect{r},\omega)\right]
 \ten{G}(\vect{r},\vect{r}',\omega)
 =\bm{\delta}(\vect{r}-\vect{r}')
\end{equation}
with $\ten{G}(\vect{r},\vect{r}',\omega)\to \tenszero$
for $|\vect{r}-\vect{r}'|\to\infty$, where the above definitions imply
\begin{equation}
\label{5}
\sum_{\lambda={e},{m}}\int\dif^3 s\,
 \ten{G}_\lambda(\vect{r},\vect{s},\omega)\sprod
 \ten{G}^{\ast\trans}_\lambda\!(\vect{r}',\vect{s},\omega)
=\frac{\hbar\mu_0}{\pi}\,\omega^2\mathrm{Im}
 \ten{G}(\vect{r},\vect{r}',\omega).
\end{equation}

For a nonuniform temperature $T\!=\!T(\vect{r})$ the thermal
state of the environment is described by $\hat{\rho}_T%
\!=\!\exp\{-\int\dif^3r\,\hat{H}_\mathrm{F}(\vect{r})%
 /[k_\mathrm{B}T(\vect{r})]\}%
/\trace(\exp\{-\int\dif^3r\,\hat{H}_\mathrm{F}(\vect{r})%
 /[k_\mathrm{B}T(\vect{r})]\})$
[$\hat{H}_\mathrm{F}\!=\!\int\dif^3r\,\hat{H}_\mathrm{F}(\vect{r})$,
$k_\mathrm{B}$: Boltzmann constant]. The relevant nonvanishing
thermal averages of the dynamical variables are thus given by
\begin{eqnarray}
\label{7}
\bigl\langle\hat{\vect{f}}_\lambda^\dagger(\vect{r},\omega)
\hat{\vect{f}}_{\lambda'}(\vect{r}',\omega')\bigr\rangle
=n_T(\vect{r},\omega)\delta_{\lambda\lambda'}
 \bm{\delta}(\vect{r}-\vect{r}')\delta(\omega-\omega'),\\
\label{8}
\bigl\langle\hat{\vect{f}}_\lambda(\vect{r},\omega)
\hat{\vect{f}}_{\lambda'}^\dagger(\vect{r}',\omega')\bigr\rangle
=[n_T(\vect{r},\omega)+1]\delta_{\lambda\lambda'}
 \bm{\delta}(\vect{r}-\vect{r}')\delta(\omega-\omega')
\end{eqnarray}
where $n_T(\vect{r},\omega)\!=\!1/
(\exp\{\hbar\omega/[k_\mathrm{B}T(\vect{r})]\}-1)$
is the average thermal photon number. 


\section{The thermal Casimir--Polder force}
\label{sec3}

The thermal CP force on an atom prepared in an incoherent
superposition of internal energy-eigenstates can in electric dipole
approximation be found from the average Lorentz force
\cite{0696,0008}
$\vect{F}(\vect{r}_{\!A},t)%
\!=\!\bigl\langle\bigl[\vect{\nabla}\hat{\vect{d}}\sprod
 \hat{\vect{E}}(\vect{r})\bigr]_{\vect{r}=\vect{r}_{\!A}}
 \bigr\rangle$. In order to evaluate this expression, one needs to
solve the coupled atom--field dynamics. Using the Hamiltonian given in
Sec.~\ref{sec2}, one finds
\begin{eqnarray}
\label{10}
\,\dot{\!\hat{A}}_{mn}=\mi\omega_{mn}\hat{A}_{mn}
 +\frac{\mi}{\hbar}\sum_k
 \bigl(\vect{d}_{nk}\hat{A}_{mk}-\vect{d}_{km}\hat{A}_{kn}\bigr)\sprod
 \hat{\vect{E}}(\vect{r}_{\!A}),\\
\label{11}
\,\dot{\!\hat{\vect{f}}}_\lambda(\vect{r},\omega)
 =-\mi\omega\hat{\vect{f}}_\lambda(\vect{r},\omega)
 +\frac{\mi}{\hbar}\sum_{m,n}\vect{d}_{mn}\sprod
\ten{G}_\lambda^\ast(\vect{r}_{A},\vect{r},\omega)\hat{A}_{mn}.
\end{eqnarray}
We eliminate the field by formally integrating the second of these
equations and substituting the result back into the first one, which
we arrange in normal ordering. After invoking the integral
relation~(\ref{5}), one obtains
\begin{eqnarray}
\label{12}
\fl
\,\dot{\!\hat{A}}_{mn}(t)
 =\mi\omega_{mn} \hat{A}_{mn}(t)
 +\frac{\mi}{\hbar}\sum_k\int_0^\infty\dif\omega
 \bigl\{\me^{\mi\omega t}
 \underline{\hat{\vect{E}}}{}^\dagger(\vect{r}_{\!A},\omega)
 \sprod\bigl[
 \vect{d}_{nk}\hat{A}_{mk}(t)-\vect{d}_{km}\hat{A}_{kn}(t)\bigr]
 \nonumber\\
 +\me^{-\mi\omega t}\bigl[\hat{A}_{mk}(t)\vect{d}_{nk}
 -\hat{A}_{kn}(t)\vect{d}_{km}\bigr]\sprod
 \underline{\hat{\vect{E}}}(\vect{r}_{\!A},\omega)\bigr\}
 +\hat{Z}_{mn}(t),
\end{eqnarray}
with 
\begin{eqnarray}
\label{13}
\fl
\hat{Z}_{mn}(t)
=\frac{\mu_0}{\hbar\pi}
 \sum_{k,l,j}\int_0^\infty\dif\omega\,\omega^2
 \vect{d}_{kl}\sprod\mathrm{Im}
 \ten{G}(\vect{r}_{\!A},\vect{r}_{\!A},\omega)\sprod
 \int_0^t\dif\tau
\bigl\{\me^{-\mi\omega(t-\tau)}\bigl[
 \vect{d}_{jm}\hat{A}_{jn}(t)
\nonumber\\
 -\vect{d}_{nj}\hat{A}_{mj}(t)\bigr]
 \hat{A}_{kl}(\tau)
 +\me^{\mi\omega(t-\tau)}\hat{A}_{kl}(\tau)\bigl[
 \vect{d}_{nj}\hat{A}_{mj}(t)-\vect{d}_{jm}\hat{A}_{nj}(t)\bigr]
 \bigr\}
\end{eqnarray}
denoting the zero-point contribution to the internal atomic dynamics.
The thermal contribution can be determined iteratively by substituting
the self-consistent solution
\begin{eqnarray}
\label{14}
\fl
\hat{A}_{mn}(t)=\me^{\mi\tilde{\omega}_{mn}t}\hat{A}_{mn}
 +\frac{\mi}{\hbar}\sum_k\int_0^\infty\dif\omega
 \int_0^t\dif\tau\,\me^{\mi\tilde{\omega}_{mn}(t-\tau)}
 \nonumber\\
\times\bigl[\hat{A}_{mk}(\tau)\vect{d}_{nk}
 -\hat{A}_{kn}(\tau)\vect{d}_{km}\bigr]\sprod
\bigl[\underline{\hat{\vect{E}}}(\vect{r}_{\!A},\omega)
 \me^{-\mi\omega\tau}+\mathrm{H.c.}\bigr]
\end{eqnarray}
to the truncated Eq.~(\ref{12}) [without $\hat{Z}_{mn}(t)$] back into
Eq.~(\ref{12}) and taking thermal expectation values with the aid of 
Eqs.~(\ref{7}) and (\ref{8}). This leads to a closed system of
equations $\bigl\langle\dot{\hat{A}}_{mn}\bigl\rangle%
=\mi\omega_{mn}\bigl\langle\hat{A}_{mn}\bigl\rangle%
+\bigl\langle\hat{Z}_{mn}\bigl\rangle%
+\bigl\langle\hat{T}_{mn}\bigl\rangle$ with a thermal contribution
\begin{eqnarray}
\label{15b}
\fl
\bigl\langle\hat{T}_{mn}(t)\bigl\rangle
=-\frac{1}{\hbar^2}
 \sum_{k,l}\int_0^\infty\dif\omega\sum_{\lambda={e},{m}}\int\dif^3 r\,
 n_T(\vect{r},\omega)\int_0^t\dif\tau
 \bigl[\me^{-\mi\omega(t-\tau)}+\me^{\mi\omega(t-\tau)}\bigr]\\
\times\bigl\{\me^{\mi\tilde{\omega}_{mk}(t-\tau)}\vect{d}_{nk}\sprod
\ten{G}_\lambda(\vect{r}_{\!A},\vect{r},\omega)\sprod
 \ten{G}^{\ast\trans}_\lambda\!(\vect{r}_{\!A},\vect{r},\omega)
 \sprod\bigl[\vect{d}_{kl}\bigl\langle\hat{A}_{ml}(\tau)\bigl\rangle
-\vect{d}_{lm}\bigl\langle\hat{A}_{lk}(\tau)\bigl\rangle
 \bigr]
 \nonumber\\
-\me^{\mi\tilde{\omega}_{kn}(t-\tau)}
 \vect{d}_{km}\sprod
\ten{G}_\lambda(\vect{r}_{\!A},\vect{r},\omega)\sprod
 \ten{G}^{\ast\trans}_\lambda\!(\vect{r}_{\!A},\vect{r},\omega)
 \sprod\bigl[\vect{d}_{nl}\bigl\langle\hat{A}_{kl}(\tau)\bigl\rangle
 -\vect{d}_{lk}\bigl\langle\hat{A}_{ln}(\tau)\bigl\rangle
 \bigr]\bigr\}.\nonumber
\end{eqnarray}

Assuming the atom--field coupling to be sufficiently weak, we can
apply the Markov approximation by writing
$\bigl\langle\hat{A}_{mn}(\tau)\bigr\rangle\!\simeq%
\!\me^{-\mi\tilde{\omega}_{mn}(t-\tau)}%
\bigl\langle\hat{A}_{mn}(t)\bigr\rangle$ and letting the lower limit
of the time integrals tend to minus infinity, so that
$\int_0^t\dif\tau\,\me^{\mi x(t-\tau)}\!\simeq\!\pi\delta(x)%
\!+\!\mi\mathcal{P}/x$. For a nondegenerate system, the off-diagonal
elements of the (internal) atomic density matrix $\hat{\sigma}$
decouple from each other as well as from the diagonal ones, and one
finds that internal atomic dynamics follows the rate equations
\begin{eqnarray}
\label{16}
\dot{\sigma}_{nn}(t)
&\!=&\!-\Gamma_n\sigma_{nn}(t)
 +\sum_k\Gamma_{kn}\sigma_{kk}(t),\\
\label{17}
\dot{\sigma}_{mn}(t)
&\!=&\!\bigl[-\mi\tilde{\omega}_{mn}
 -{\textstyle\frac{1}{2}}(\Gamma_m+\Gamma_n)/2\bigr]
 \sigma_{mn}(t)\qquad\mbox{for }m\neq n
\end{eqnarray}
($\sigma_{mn}$ $\!=$ $\!\langle m|\hat{\sigma}|n\rangle$ $\!=$
$\!\bigl\langle\hat{A}_{nm}\bigr\rangle$). As follows from
Eqs.~(\ref{13}) and (\ref{15b}), the total loss rates read 
\begin{eqnarray}
\label{18}
\fl
\Gamma_n=\sum_k\Gamma_{nk}
 =\frac{2\mu_0}{\hbar}\sum_k\tilde{\omega}_{nk}^2
 \Theta(\tilde{\omega}_{nk})\vect{d}_{nk}\sprod\mathrm{Im}
 \ten{G}(\vect{r}_{\!A},\vect{r}_{\!A},|\tilde{\omega}_{nk}|)
 \sprod\vect{d}_{kn}\nonumber\\
+\,\frac{2\pi}{\hbar^2}\sum_k\sum_{\lambda={e},{m}}\int\dif^3 r\,
 [\Theta(\tilde{\omega}_{nk})n_T(\vect{r},\tilde{\omega}_{nk})
 +\Theta(\tilde{\omega}_{kn})n_T(\vect{r},\tilde{\omega}_{kn})]
 \nonumber\\
\qquad\times\vect{d}_{nk}\sprod
 \ten{G}_\lambda(\vect{r}_{\!A},\vect{r},|\tilde{\omega}_{nk}|)\sprod
 \ten{G}^{\ast\trans}_\lambda
 (\vect{r}_{\!A},\vect{r},|\tilde{\omega}_{nk}|)\sprod\vect{d}_{kn}
\end{eqnarray}
and the shifts of atomic transition frequencies
$\tilde{\omega}_{mn}\!=\!\omega_{mn}%
\!+\!\delta\omega_m\!-\!\delta\omega_n$ 
are given by
\begin{eqnarray} 
\label{19}
\fl
\delta\omega_n=\sum_k \delta\omega_{nk}
=\frac{\mu_0}{\pi\hbar}\sum_k
 \mathcal{P}\int_0^\infty\dif\omega\,\omega^2
\frac{\vect{d}_{nk}\sprod\mathrm{Im}
 \ten{G}^{(1)}(\vect{r}_{\!A},\vect{r}_{\!A},\omega)\sprod
 \vect{d}_{kn}}{\tilde{\omega}_{nk}-\omega}
 \nonumber\\
+\frac{1}{\hbar^2}\sum_k
 \mathcal{P}\int_0^\infty\dif\omega
 \sum_{\lambda={e},{m}}\int\dif^3 r\,
 \biggl[\frac{n_T(\vect{r},\omega)}{\tilde{\omega}_{nk}-\omega}
+\frac{n_T(\vect{r},\omega)}{\tilde{\omega}_{nk}+\omega}\biggr]
 \nonumber\\
\qquad\times\vect{d}_{nk}\sprod
 \ten{G}_\lambda(\vect{r}_{\!A},\vect{r},\omega)
 \sprod\ten{G}^{\ast\trans}_\lambda(\vect{r}_{\!A},\vect{r},\omega)
 \sprod\vect{d}_{kn}.
\end{eqnarray}
Note that we have replaced the Green tensor with its scattering part
$\ten{G}^{(1)}$ for the zero-point frequency shift since the
free-space zero-point Lamb shift is thought to be already included in
the bare transition frequencies $\omega_{mn}$.

Having solved the atom--field dynamics, we can now evaluate the
average Lorentz force by using Eqs.~(\ref{1}), (\ref{7}), (\ref{8}), 
(\ref{14}) and the solution to Eq.~(\ref{11}). After invoking
the correlation function
$\bigl\langle\hat{A}_{mn}(t)\hat{A}_{kl}(\tau)\bigr\rangle%
\!=\!\delta_{nk}\bigl\langle\hat{A}_{ml}(\tau)\bigr\rangle
 \me^{\mi\Omega_{mn}(t-\tau)}$
[$\Omega_{mn}\!=\!\tilde{\omega}_{mn}%
\!+\!\mi(\Gamma_m\!+\!\Gamma_n)/2$]
which follows from Eq.~(\ref{17}) via the quantum regression theorem
\cite{0605}, one obtains 
\begin{eqnarray} 
\label{19a}
\fl
\vect{F}(\vect{r}_{\!A},t)
 =\frac{\mi}{\hbar}\sum_{n,k}\int_0^\infty\dif\omega
 \sum_{\lambda={e},{m}}\int\dif^3 s\,
\vect{\nabla}\tprod\vect{d}_{nk}\sprod
\ten{G}_\lambda(\vect{r},\vect{s},\omega)
 \sprod\ten{G}^{\ast\trans}_\lambda(\vect{r}_{\!A},\vect{s},\omega)
\sprod\vect{d}_{kn}\bigr|_{\vect{r}=\vect{r}_{\!A}} 
\nonumber\\
\fl\quad
\times\int_0^t\dif\tau\,\bigl\langle\hat{A}_{nn}(\tau)\bigl\rangle
 \bigl\{n_T(\vect{s},\omega)
 \me^{\mi(\omega+\Omega_{nk})(t-\tau)}
 +[n_T(\vect{s},\omega)+1]
 \me^{-\mi(\omega-\Omega_{nk})(t-\tau)}\bigr\}
 +\mathrm{c.c.}
\end{eqnarray}
Using the Markov approximation in the form 
$\int_0^t\dif\tau\,\bigl\langle\hat{A}_{nn}(\tau)\bigl\rangle\ldots%
\!\simeq\!\bigl\langle\hat{A}_{nn}(t)\bigl\rangle%
\int_{-\infty}^t\dif\tau\,\ldots$, the thermal CP force for an atom
prepared in an incoherent superposition of internal energy-eigenstates
is given by $\vect{F}(\vect{r}_{\!A},t)%
\!=\!\sum_{n}\sigma_{nn}(t)\vect{F}_n(\vect{r}_{\!A})$ with force
components
\begin{eqnarray} 
\label{20}
\fl
\vect{F}_n(\vect{r}_{\!A})
 =\frac{1}{\hbar}\sum_{k}\int_0^\infty\dif\omega
 \sum_{\lambda={e},{m}}\int\dif^3s\,
 \biggl\{\frac{n_T(\vect{s},\omega)+1}{\omega-\Omega_{nk}}
 -\frac{n_T(\vect{s},\omega)}{\omega+\Omega_{nk}}\biggr\}\nonumber\\
\times\vect{\nabla}\tprod\vect{d}_{nk}\sprod
\ten{G}_\lambda(\vect{r},\vect{s},\omega)
 \sprod\ten{G}^{\ast\trans}_\lambda(\vect{r}_{\!A},\vect{s},\omega)
\sprod\vect{d}_{kn}\bigr|_{\vect{r}=\vect{r}_{\!A}} 
 +\mathrm{c.c.}
\end{eqnarray}

An important special case is that of an equilibrium environment at
temperature $T$. Applying Eq.~(\ref{5}) to the $\vect{s}$-integral in
Eq.~(\ref{20}) and transforming the $\omega$-integral by means of
contour-integral techniques, one obtains in the perturbative limit
($\Omega_{nk}\!\simeq\!\omega_{nk}$) 
\begin{eqnarray}
\label{21}
\fl
\vect{F}_n(\vect{r}_{\!A})
 =-\mu_0k_\mathrm{B}T\sum_{N=0}^\infty
 \bigl(1-{\textstyle\frac{1}{2}}\delta_{N0}\bigr)\xi_N^2
 \vect{\nabla}_{\!\!A}\trace\bigl[\bm{\alpha}_n(\mi\xi)
 \sprod\ten{G}^{(1)}(\vect{r}_{\!A},\vect{r}_{\!A},\mi\xi_N)
 \bigr]\\
\fl\qquad+\mu_0\sum_k\bigl\{\Theta(\omega_{nk})[n(\omega_{nk})+1]
 -\Theta(\omega_{kn})n(\omega_{kn})\bigr\}
 \omega^2_{nk}\vect{\nabla}_{\!\!A}\vect{d}_{nk}\sprod
 \mathrm{Re}\ten{G}^{(1)}(\vect{r}_{\!A},\vect{r}_{\!A},\Omega_{nk})
 \sprod\vect{d}_{kn}\nonumber
\end{eqnarray}
($\xi_N=2\pi k_\mathrm{B}TN/\hbar$, Matsubara frequencies), where the
atomic polarisability reads
\begin{equation}
\label{Eq21}
\bm{\alpha}_n(\omega)
=\lim_{\epsilon\to 0}\frac{1}{\hbar}\sum_k\biggl[
 \frac{\vect{d}_{nk}\tprod\vect{d}_{kn}}
 {-\omega_{nk}-\omega-\mi\epsilon}
 +\frac{\vect{d}_{kn}\tprod\vect{d}_{nk}}
 {-\omega_{nk}+\omega+\mi\epsilon}\biggr].
\end{equation}
This is in agreement with previous results \cite{0824}.
Equations~(\ref{16})--(\ref{18}) show that in the
long-time limit one has $\hat{\sigma}(t\to\infty)\!=\!\hat{\sigma}_T%
\!=\!\exp[-\hat{H}_A/(k_\mathrm{B}T)]/%
\trace\exp[-\hat{H}_A/(k_\mathrm{B}T)]$, and the 
Lifshitz force $\vect{F}(\vect{r}_{\!A},t\to\infty)%
\!=\!-\mu_0k_\mathrm{B}T\sum_{N=0}^\infty%
\bigl(1-{\textstyle\frac{1}{2}}\delta_{N0}\bigr)\xi_N^2%
\alpha_T(\mi\xi)\vect{\nabla}_{\!\!A}\trace%
\ten{G}^{(1)}(\vect{r}_{\!A},\vect{r}_{\!A},\mi\xi_N)$ is recovered
with a thermal polarisability
$\alpha_T(\omega)\!=\!\sum_n\sigma_{T,nn}\alpha_n(\omega)$.


\section{Summary}
\label{sec4}

We have used macroscopic QED to obtain a general expression for the
CP force on an atom which is placed in an arbitrary magnetoelectric
environment of nonuniform temperature and whose initial internal
state may be an arbitrary superposition of internal
energy-eigenstates. Our result reduces to previous ones in the special
case of uniform temperature.


\ack

This work was supported by the Alexander von Humboldt Foundation and
the UK Engineering and Physical Sciences Research Council.


\end{document}